\documentclass[12pt]{article}
\usepackage{a4wide}
\usepackage{graphicx}
\usepackage{amssymb}
\usepackage{amsmath}
\usepackage{authblk}
\usepackage{cite}
\usepackage{slashed}
\usepackage{braket}
\newcommand{\be}{\begin{equation}}
\newcommand{\ee}{\end{equation}}
\newcommand{\ba}{\begin{eqnarray}}
\newcommand{\ea}{\end{eqnarray}}

\date{}
\title{Two-portal Dark Matter}

\vspace{1.cm}

\author{Karim Ghorbani \thanks{kghorbani@ipm.ir}}   
\affil {\small\it{Physics Department, Faculty of Sciences, Arak University,  \it Arak 38156-8-8349, Iran}}

\author{Hossein Ghorbani\thanks{pghorbani@ipm.ir}}
\affil {\small\it{Institute for Research in Fundamental Sciences (IPM)\\
 \it School of Particles and Accelerators,  \it P.O. Box 19395-5531, Tehran, Iran}}

\begin{document}
 
\maketitle

\abstract{
We propose a renormalizable dark matter model in which a fermionic dark matter (DM) candidate 
communicates with the standard model
particles through two distinct portals: Higgs and vector portals. The dark sector is charged under a $U(1)'$ 
gauge symmetry while the standard model has a leptophobic interaction with the dark vector boson. 
The leading contribution of DM-nucleon elastic scattering cross section begins at one-loop level. 
The model meets all the constraints imposed by direct detection experiments provided by LUX 
and XENON100, observed relic abundance according to WMAP and Planck, and the invisible Higgs decay width 
measured at the LHC. It turns out that the dark matter mass in the viable parameter space can take values 
from a few GeV up to 1 TeV. 
This is a new feature which is absent in the models with only one portal.
In addition, we can find in the constrained regions of the parameter space a DM mass 
of $\sim 34$ GeV annihilating into $b$ quark pair, which 
explains the Fermi-LAT gamma-ray excess.
}
\newpage

\section{Introduction}
\label{sect:introduction}
Cosmological observations indicate unequivocally that the  
Standard Model (SM) particles constitute only 5\% of the mass  
content of our Universe, the rest 26\% dark matter (DM) and 69\%
dark energy are yet unknown \cite{Ade:2013zuv,Hinshaw:2012aka}. 
If the weakly interacting massive particle (WIMP) is a correct scenario for DM, 
a big question however, is what the dark matter is made of and what would be 
the fundamental interaction of its constituents with the ordinary matter. 
Direct detection experiments are designed to probe dark matter (DM) elastic scattering off 
nuclei. In this regards, underground LUX \cite{Akerib:LUX} and XENON100 \cite{Aprile:XENON100} 
dark matter experiments so far have found no signal on these type of interactions, even though 
they provide us with an upper limit on the elastic scattering cross section. 

Recent observation of the galactic center gamma-ray excess (GCE), given its intensity and 
spatial morphology, can be explained by the dark matter annihilation in the galactic center (GC) 
\cite{Hooper:2010mq,Abazajian:2012pn,Gordon:2013vta,Abazajian:2014fta,Daylan:2014rsa,Calore:2014xka,Calore:2014nla}. 
There is a large number of models in the literature suggested in order to explain 
the gamma-ray excess but among them are models with dark matter, annihilating 
predominantly into $b$ quark pair and on top of that can evade stringent 
bounds from direct detection experiments 
\cite{Boehm:2014hva,Ipek:2014gua,Izaguirre:2014vva,Cheung:2014lqa,Ghorbani:2014qpa,Dolan:2014ska,Gherghetta:2015ysa}.

Moreover, there are models as extensions to the SM
with an additional $U(1)'$ gauge symmetry and its associated $Z'$ boson which 
contain a DM candidate annihilating via an intermediate neutral gauge boson 
$Z^\prime$ 
\cite{Dudas:2009uq,Chu:2013jja,Arcadi:2013qia,Lebedev:2014bba,Belanger:2007dx,Buckley:2011mm,Pospelov:2007mp,Hooper:2014fda,Berlin:2014tja,Mizukoshi:2010ky,Alves:2015pea}.

Models with an extra broken $U(1)'$ gauge symmetry are motivated by 
new physics beyond the SM, as examples for various models 
we recall, those focusing on the grand unified theories like $SO(10)$ and $E_6$,  
(see e.g. \cite{London:1986,Hewett:1988xc} and for a review consult \cite{Langacker:2008yv}), 
dynamical symmetry breaking models like topcolor (see \cite{Hill:2002ap} for a review),
decoupled models like leptophobic $Z'$ \cite{delAguila:1986iw}, Little Higgs theories 
\cite{ArkaniHamed:2001is,ArkaniHamed:2002qx,Han:2003wu,Perelstein:2005ka}, Twin Higgs 
model \cite{Chacko:2005pe}, family non-universal scenario \cite{Demir:2005ti}.

There are many models where the SM and DM sectors interact through only 
one portal, in the sense that there is only one type 
of mediator (e.g. scalar or vector) to connect the two sectors. There are also models with two 
similar mediators \cite{Ghorbani:2014gka, Cline:2015qha}. 

In this work we propose for the first time a {\it minimal} dark 
matter model with two distinct portals in tandem: Higgs and vector ones. 
In this article we construct a {\it two-portal} DM model based on an extra 
gauge symmetry $U(1)'$ which not only can explain the observed relic density and the 
galactic gamma-ray excess but also can evade direct detection as well as the 
constraints from invisible Higgs decay.

In our two-portal model the dark sector consists of a Dirac fermion as
a WIMP dark matter candidate, 
a complex scalar field as the first mediator, both of them  
charged under a new $U(1)^\prime$ gauge symmetry. Obviously the fermion dark matter is
coupled to the dark gauge boson $Z'$, covariantly. In addition, we assume that only the SM quarks
are also charged under the $U(1)^\prime$. Thus, the $Z'$ field interacts with the SM particles and the DM,
hence the second mediator of the model.
In the present model we will deal with a non-universal $Z'$ gauge boson, such 
that the new gauge boson has negligible coupling to the first and second generation
of the SM quarks. 

Models with a new $Z'$ gauge boson which prefer interaction with only the third 
family of the SM fermions are widely investigated within different 
scenarios beyond the SM. 
Among these scenarios, we recall superstring inspired models \cite{delAguila:1986iw}, 
topcolor assisted technicolor model \cite{Hill:1994hp}, 
phenomenology of a $Z'$ boson coupled only to third-family fermions \cite{Andrianov:1998hx}, 
electroweak constraints on models with non-universal $Z'$ bosons \cite{Chivukula:2002ry}, and 
warped models in which the extra $Z'$ boson typically couples only to the third generation \cite{Belanger:2007dx}.     

The present article has the following structure. In the next section we introduce 
our model in detail. In Sec.~\ref{invisible-constrain} we compute the SM Higgs invisible decay 
width within the model and address the constraints on the invisible decay width from 
the LHC measurements.              
We derive a formula for the DM-nucleon cross section in Sec.~\ref{direct}.
In Sec.~\ref{relic-density} we discuss on the DM relic density 
within the thermal freeze-out mechanism
and our numerical computations for the relic 
abundance and DM-nucleon interaction are discussed in Sec.~\ref{Numerical}.
In Sec.~\ref{gamma-ray-emission} we will find regions in the viable parameter space of the two-portal model  
that can explain the Galactic gamma-ray excess given the recent Fermi-LAT data analysis.
We finally finish with a conclusion.

\section{The Model} 
\label{model}
We introduce a dark matter model which has the property of having
DM-SM interaction through two different portals, i.e. the vector portal and the
Higgs portal. We will show in Sec.~\ref{direct} that for the model to be elaborated below there is no tree level DM 
scattering off nuclei and the Feynman diagrams begin with a one-loop contribution which turns out to be suppressive. 
A two-portal dark matter model can therefore be designed in order to predict the DM elastic scattering 
to be consistent with the direct detection experiments. 
This is reminiscent to the velocity suppressed models for elastic scattering processes but in a different way.  

The details of the model comes in the following: beside having a scalar field that mixes with the SM
Higgs via the Higgs portal, to have a vector portal interaction, we
assume a $U(1)'$ gauge theory in the dark sector as the simplest model including
a gauge boson. We also assume that only the SM quarks (but not the leptons) are charged 
under the $U(1)'$. In other words, we are dealing with a {\it leptophobic}
vector portal interaction.

The total Lagrangian consists of
the standard model part, the dark sector and the interactions between these two sectors: 
\ba
\mathcal{L}=\mathcal{L}_{\text{SM}}+\mathcal{L}_{\text{DM}}+\mathcal{L}_{\text{int}} \,.
\ea
The SM covariant derivative acting on the quarks must now be modified as
\ba
 D^{\text{SM}}_\mu \rightarrow D'^{\text{SM}}_\mu=D^{\text{SM}}_\mu -i g' \frac{z}{2} Z'_\mu \,,
\ea
where $z$ is the dark charge of the quark field that the covariant derivative acts on. 

The dark matter Lagrangian consists of a fermionic dark matter and a complex scalar field both charged under  
$U(1)'$:

\ba 
\label{LDM}
 \mathcal{L}_{\text{DM}}=-\frac{1}{4} F'_{\mu\nu}F'^{\mu\nu}+\bar{\chi}\left(i\gamma^{\mu}D'_{\mu}-m_{\chi}\right)\chi
+\left(D'_{\mu}\phi\right)\left(D'^{\mu}\phi\right)^*-m_{\phi}^{2}(\phi \phi^*)
-\frac{1}{4}\lambda (\phi \phi^*)^2 \,,
\ea
where $F'^{\mu\nu}$ is the $U(1)'$ field strength, $\chi$ is a Dirac fermion as the dark matter candidate and $\phi$
is a complex scalar field. 
The dark sector covariant derivative is given by 
\ba
D'_{\mu}=\partial_{\mu}- i g' \frac{z}{2} Z'_\mu.
\ea
We assume that the interaction of the standard model particles with the $U(1)'$ 
gauge boson, $Z'$ is leptophobic. 
In other words, non of the leptons in the SM are charged under $U(1)'$. 
The $\mathcal{L}_{\text{int}}$ consisting of the scalar-Higgs and $Z'$-quark interactions reads: 
\ba
\label{Lint}
\mathcal{L}_{\text{int}}=-\lambda(\phi \phi^*) \left(HH^{\dagger}\right)
+ g' \frac{z_{Q_L}}{2} Z'_{\mu}\bar{Q}_L\gamma^{\mu}Q_L 
+ g' \frac{z_{u_R}}{2} Z'_{\mu}\bar{u}_R\gamma^{\mu}u_R
+ g' \frac{z_{d_R}}{2} Z'_{\mu}\bar{d}_R\gamma^{\mu}d_R\,,
\ea
where $H$ is the SM Higgs doublet, $Q_L$, $u_R$ and $d_R$ are respectively 
left-handed quark doublet, right-handed up-quark singlet
and right-handed down-quark singlet. 
The couplings of the light quarks $u,d,...$ with $Z'$ are considered to be negligible.
Therefore, by $z_{Q_L}$, $z_{u_R}$ and $z_{d_R}$ we mean the dark charge of only the third quark family
i.e. the $t$ and $b$ quarks. The upshot is that the dark 
matter scattering off nuclei lacks the tree level contribution (see Fig.~\ref{direct}) 
and remains suppressed as expected from direct detection experiments.

Having introduced dark gauge boson, $Z'$, interacting with SM and DM fermionic currents, 
one should note that new anomalies from triangle Feynman diagrams may arise. 
However, it can be shown that assigning appropriate dark charges for quarks can lead to a anomaly-free theory 
\footnote{For some details on anomaly-free conditions in the extended SM including $U(1)'$ interactions 
with no additional fermions see \cite{Liu:2011dh}. In our model we have an additional fermion that is the dark matter
Dirac field. Taking equal dark charges for left- and right-handed components of the Dirac 
fermion $z_{\chi_L}=z_{\chi_R}$ we lead to the same anomaly-free conditions mentioned in \cite{Liu:2011dh}.  
}.

The anomaly-free conditions put constraints on the top and bottom quark $U(1)'$ charges: 
$z_{Q_L}=-2$, $z_{u_R}=+2$ and $z_{d_R}=+2$.
Substituting these charges in Eq. (\ref{Lint}) the $\mathcal{L}_{\text{int}}$ becomes:
\ba
 \mathcal{L}_{\text{int}}=-\lambda_{1} (\phi \phi^*) \left(HH^{\dagger}\right)
+ g' Z'_{\mu}\bar{t}\gamma^{\mu}\gamma^5t 
+ g' Z'_{\mu}\bar{b}\gamma^{\mu}\gamma^5b.
\ea

The $\mathcal{L}_{\text{int}}$ consists of 
the Higgs portal where the scalar field interacts with the SM Higgs quadratically, and the vector portal where 
the dark gauge boson interacts axially with the third family quarks. 
 
\begin{figure}
\begin{center}
\includegraphics[scale=.2,angle =0]{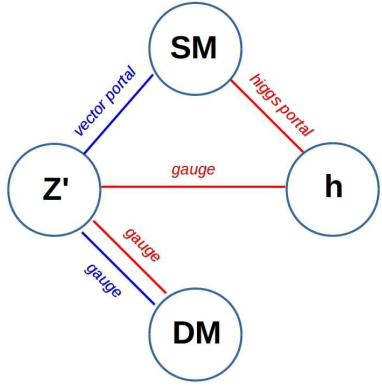}
\end{center}
\caption{ Two-portal SM-DM interactions: the dark matter candidate indirectly can 
interact with standard model through a Higgs and a vector portal. In this model the 
dark matter can interact directly with only one of the mediators i.e. the dark vector boson.}
\label{modeldiagram}
\end{figure}

The scalar field $\phi$ does not interact directly with the DM particle
$\chi$ but interacts with that only through another mediator of the model
i.e. the dark gauge boson $Z'$. On the other
hand, the gauge boson mediator interacts directly with the dark matter particle which 
can be seen by following the red line 
in Fig.~\ref{modeldiagram}. 
The scalar field $\phi$ and the gauge boson $Z'$ are connected
to the SM respectively via the Higgs portal and
via an interaction with the third family of quarks (vector portal). The novelty of the current minimal model is
that there are two distinct mediators at the same time which makes a bridge
between the DM sector and the SM sector. Schematically these interactions
are shown in Fig.~\ref{modeldiagram}. 

The SM Higgs potential is given by 
\ba
V_{H}= - \mu_{H}\left(HH^{\dagger}\right) - \lambda_{H}\left(HH^{\dagger}\right)^{2}\,,
\ea
where the Higgs doublet takes on a non-zero vacuum expectation value (vev), 
\ba
\label{vevhiggs}
H=\frac{1}{\sqrt{2}}\left(\begin{array}{c}
0\\
v+\tilde{h}
\end{array}\right).
\ea
 We assume that the scalar mediator also takes a non-zero vev,
\ba
\label{vevphi}
 \braket\phi  = v^\prime\Rightarrow\phi=v'+\frac{1}{\sqrt{2}}\tilde{h}'.
\ea
$\tilde h$ and $\tilde h'$ are respectively the SM Higgs and the singlet scalar field 
fluctuations around their vacuum expectation values.
It is worth mentioning that once the complex scalar develops a non-zero vacuum expectation value, $v'$, this
breaks the $U(1)'$ symmetry spontaneously and the $Z'$ boson will acquire mass. 
On top of that, the strength of these vertices, $Z' Z' h$ and $Z' Z' h'$, 
is proportional to  $v'$. Therefore, according to the Feynman diagram in Fig.~\ref{dark-direct}, 
the choice  $v' = 0$ will give rise to zero DM-quark elastic scattering cross section which 
is a trivial scenario.

After substituting Eq. (\ref{vevphi}) in Eq. (\ref{LDM}) 
and expanding the Lagrangian, the mass of the 
dark gauge boson turns out to be $g' v'/\sqrt{2}$.  
As may be followed in \cite{Ghorbani:2014qpa} the masses of the SM Higgs
particle $h$ and the scalar mediator $h'$ can be obtained by diagonalizing
the mass matrix, 
\ba
\label{matrix}
M = \bordermatrix{~ & \tilde{h} & \tilde{h}' \cr
                  \tilde{h} & 2\lambda_{H}v^{2} & \sqrt{2} \lambda_{1} v v^\prime \cr
                  \tilde{h}' & \sqrt{2} \lambda_{1} v v^\prime & \frac{1}{2}\lambda {v^\prime}^{2}-\frac{1}{2}\lambda_{1} v^{2} \cr}\,,
                  \ea
where we have used the following relations coming from minimizing
the total potential, 
\ba
 m_{\phi}^{2}=-\lambda{v^\prime}^{2}-\lambda_{1} v^{2}\,,
\ea
\ba
\mu_{H}^{2}=-\lambda_{H}v^{2}-\lambda_{1} {v^\prime}^{2}. 
\ea
We can redefine the scalars $\tilde{h}$ and $\tilde{h}'$ by introducing
a mass mixing angle in order to get a diagonalized mass matrix, 
\ba
 h=\sin\left(\theta\right)\tilde{h}+\cos\left(\theta\right)\tilde{h}'\,,
\ea
\ba
 h'=\cos\left(\theta\right)\tilde{h}-\sin\left(\theta\right)\tilde{h}'\,,
\ea
with the mixing angle $\theta$ being
\ba
\label{mixing-angle}
 \tan\left(\theta\right)=\frac{1}{1+\sqrt{1+y^{2}}}\,,
 \,\,\,\,\,\,\,\,\,\,\,\,\,\,\,\,\,\,\, y=\frac{2m_{\tilde{h}\tilde{h}'}^{2}}{m_{\tilde{h}}^{2}-m_{\tilde{h}'}^{2}}.
\ea
where $m_{\tilde{h}\tilde{h}'}$ is the off-diagonal entry of the mass matrix in Eq.~(\ref{matrix}). 

The masses of the redefined scalar fields read,
\ba
\label{mass-eigen}
m_{h}^{2},m_{h'}^{2}=\frac{m_{\tilde{h}}^{2}+m_{\tilde{h}'}^{2}}{2}\pm\frac{m_{\tilde{h}}^{2}
-m_{\tilde{h}'}^{2}}{2}\sqrt{1+y^{2}}\,, 
\ea
where the upper sign (lower sign) corresponds to $m_{h}$ ($m_{h'}$).
The standard model Higgs is denoted here by $h$ with mass $m_{h}=125$
GeV and $h^{\prime}$ is the singlet scalar.
Exploiting Eq.~(\ref{mixing-angle}) and Eq.~(\ref{mass-eigen})  
we can obtain the quartic couplings as a function of SM Higgs mass, singlet scalar mass, 
the mixing angle and the vacuum expectation values $v$ and $v'$,
\ba
\label{couplings}
\lambda_{H}  = \frac{m^{2}_{h^\prime} \sin^2 \theta +m^{2}_{h} \cos^2 \theta }{2v^{2}}\,,
\nonumber\\
\lambda  = \frac{m^{2}_{h'} \cos^2 \theta +m^{2}_{h} \sin^2 \theta }{v^2/2} 
      - \frac{v^2}{{v^\prime}^2} \lambda_{1}  \,,
\nonumber\\
\lambda_{1} = \frac{m^{2}_{h}-m^{2}_{h^\prime}}{2\sqrt{2}vv^{\prime}} \sin 2\theta \,.
\ea  
The vacuum stability of the total potential is equivalent to having positive 
eigenvalues for the scalar boson mass-squared matrix.
At tree-level this brings in the following constraints on the couplings (see also \cite{Ghorbani:2014qpa}); 
$\lambda_{H} > 0$, $\lambda {v^\prime}^2 > \lambda_{1} {v^2}$ and 
${v^\prime}^2 (\lambda_{H}\lambda -2 \lambda_{1}^2) > v^2 \lambda_{1} \lambda_{H}$.   
In our numerical analysis we will choose $m_{\chi}, m_{h^\prime}$, $\theta$, $v^\prime$ and $g^\prime$  
as free parameters. 
 
\section{Invisible Higgs Decay}
\label{invisible-constrain}

In the present model, there are two new decay channels for the SM Higgs which can 
modify the total decay width of the Higgs boson within the SM. 
The current measurement of total decay width for the 125 GeV Higgs reads, 
$\Gamma^{\text{SM}}_{\text{Higgs}} \sim 4$ MeV \cite{Heinemeyer:2013}. 
In case the dark gauge boson is light enough such that 
$m_{Z^\prime} < m_{\text{h}}/2$ the Higgs boson is kinematically allowed to undergo 
the following invisible decay 
\ba
\label{dark-decay1}
\Gamma_{\text{inv}}(h \to Z^\prime Z^\prime ) = \frac{ {v^\prime}^2 {g^\prime}^4 \sin^2\theta  }{16\pi m_{h}} 
(1-4 m^{2}_{Z^\prime}/m_{h}^2)^{1/2}.
\ea
In addition when we consider light scalar boson with $m_{h^\prime} < m_{h}/2$, 
another decay channel is plausible for the SM Higgs with
\ba
\label{dark-decay2}
\Gamma_{\text{inv}}(h \to h^\prime h^\prime ) = \frac{c^2}{128\pi m_{h}} (1-4 m^{2}_{h^\prime}/m_{h}^2)^{1/2},
\ea
where 
\ba
c = 3 \sqrt{2} \lambda v^\prime \cos^2 \theta \sin \theta 
    +12 \lambda_{H} v  \cos \theta \sin^2 \theta 
    -6 \lambda_{1} v  \cos \theta \sin^2 \theta  \\
\nonumber
   +2 \lambda_{1} v  \cos \theta  
   + 6\sqrt{2} \lambda_{1} v^\prime \sin^3 \theta 
   - 4 \sqrt{2}  \lambda_{1} v^\prime  \sin \theta. 
\ea
We thus expect the total Higgs decay width to modify as 
\ba
\label{total-width}
\Gamma^{\text{tot}}_{\text{Higgs}} = \cos^2\theta~\Gamma^{\text{SM}}_{\text{Higgs}} 
 + \Theta (m_{h}- 2 m_{Z^\prime}) \Gamma(h \to Z^\prime Z^\prime)
 + \Theta (m_{h}- 2 m_{h^\prime}) \Gamma(h \to h^\prime h^\prime),
\ea  
where $\Theta$ is the step function. 
It is worth mentioning that since here Higgs has no tree level interaction with the DM, 
the invisible Higgs decay width is independent of DM mass at leading order. 
There exist an experimental upper limit for the invisible 
branching ratio of the 125 GeV Higgs decay 
investigated at the LHC, $\text{Br}_{\text{inv}}  \lesssim 0.35$ \cite{Belanger:invisible}.       
In our numerical analysis when applicable, we restrict ourself 
into the parameter space which satisfies the condition 
$\Gamma_{\text{inv}} / \Gamma^{\text{tot}}_{\text{Higgs}}  \lesssim 0.35$.

\section{Direct Detection}
\begin{figure}
\begin{center}
\includegraphics[scale=.8,angle =0]{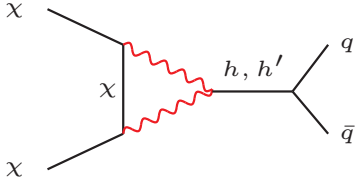}
\end{center}
\caption{The Feynman diagram for the DM elastic scattering with quarks. The wavy lines
stand for the propagation of the $Z'$ boson.}
\label{dark-direct}
\end{figure}

The tree level DM-quark elastic scattering is suppressed because 
dark vector boson interaction with light quarks are assumed to be negligible. 
As depicted in Fig.~\ref{dark-direct} the first leading contribution 
to the elastic scattering amplitude is obtained through a one-loop interaction 
coupled to the SM Higgs or the Higgs-like scalar 
where the DM particle and dark gauge boson run in the loop. The DM-quark 
scattering amplitude is obtained as
\ba
{\cal M} = -4i g'^4 v' \frac{m_{q}}{v} 
\Big[\frac{\sin^2 \theta}{(p_1-p_2)^2-m_{h}^2} -\frac{\cos^2 \theta}{(p_1-p_2)^2-m_{h'}^2}  \Big] \bar q q~  \times 
\nonumber \\
\int \frac{d^4q}{(2\pi)^2}  \frac{\bar \chi(p_{2}) \gamma_{\mu} (\slashed{q}+m_{\chi}) \gamma^{\mu} \chi(p_1) }
{[(p_2-q)^2-m_{Z^{'}}^2][(p_1-q)^2-m_{Z^{'}}^2][q^2-m_{\chi}^2]}   \,,
\ea
where respectively $p_1$ and $p_2$ are the four-momenta of the incoming and outgoing DM, 
and the DM four-momentum in the loop is denoted by $q$.  
Since $(p_1-p_2)^2 \ll m_{\chi}^2, m_{Z'}^2$, we can then perform the loop integral at $t = (p_1-p_2)^2 \sim 0$ to get the 
effective scattering amplitude\footnote{Currently, the typical recoil energy under examination at direct detection 
experiments is $E_{R} \sim 10$ KeV. On the other hand, the momentum 
transferred to a nucleus of mass $M_{N}$ is given by $t = 2 M_{N} E_{R}$. For a xenon 
nucleus for example, we obtain $t \sim 2 \times 10^{-3}$ GeV$^2$. Thus we expect $t \ll m_{\chi}^2, m_{Z'}^2$. } 
\be
{\cal M}_{\text{eff}} =  \frac{g'^4 v'}{4\pi^2m_{\chi}} S(\beta) \frac{m_{q}}{v} 
\Big[ \frac{\cos^2 \theta}{m_{h'}^2}-\frac{\sin^2 \theta}{m_{h}^2}  \Big] (\bar q q)(\bar \chi \chi) 
 \equiv  \alpha_{q}~(\bar q q)( \bar \chi \chi) \,,
\ee
where 
\ba
S(\beta) = -2 +\beta \log \beta  - \frac{\beta^2-2\beta-2}{\sqrt{\beta^2-4\beta}}
\log \frac{\sqrt{\beta} + \sqrt{\beta-4}}{\sqrt{\beta} - \sqrt{\beta-4}} \,,
\ea
and $\beta = (\frac{m_{Z^{'}}}{m_{\chi}})^2$. In order to find the DM-nucleon 
elastic scattering cross section one needs to evaluate the nucleonic 
matrix element. However, at the vanishing momentum transfer
we can make use of the conventional assumption that the nucleonic
matrix element with quark current is proportional to the nucleonic
matrix element with nucleon current \cite{Belanger:2008-Direct,Ellis:2008,Nihei:2004}
\ba
\sum_{\text{q}} \alpha_{q} \langle N_{f}|\bar q q|N_{i} \rangle\  \equiv
 \alpha_{N} \langle N_{f}|\bar N N|N_{i} \rangle\ \,, 
\ea     
in which 
\ba
\alpha_{N} = m_{N} \Big( \sum_{q = u,d,s} f^{N}_{Tq} \frac{\alpha_{q}}{m_{q}} 
+ \frac{2}{27} f^{N}_{Tg} \sum_{q = c,b,t}   \frac{\alpha_{q}}{m_{q}} \Big)  \,.
\ea 
\label{direct}
The scalar couplings $f^{N}_{Tq}$ and $f^{N}_{Tg}$ are 
responsible for the low energy strong interaction and 
nucleon mass is denoted by $m_{N}$. 
In the numerical computation in Sec.~\ref{Numerical} we shall use the following values for 
the scalar couplings \cite{Belanger:MICRO}
\ba
f^{p}_{u} = 0.0153,~~~~~~~ f^{p}_{d} = 0.0191, ~~~~~~~ f^{p}_{s} = 0.0447 \,.
\ea    
Spin-independent (SI) total cross section of DM-nucleon elastic 
scattering is finally achieved as 
\ba
\label{cross-section}
\sigma^{\text{N}}_{\text{SI}} = 
\frac{4 \alpha_{N}^2 \mu_{\chi N}^2}{\pi}\,,
\ea  
where $\mu_{\chi N}$ is the reduced mass of the DM-nucleon system.

\section{Relic Abundance}
\label{relic-density}
\begin{figure}
\begin{center}
\includegraphics[scale=.8,angle =0]{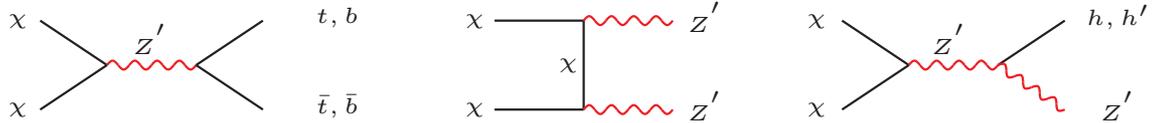}
\end{center}
\caption{The Feynman diagrams for the DM annihilation processes.}
\label{dark-ann}
\end{figure}
The fermionic dark matter candidate in the model laid out earlier is of WIMP type 
DM whose present day density, the so called relic density, is a remnant from freeze-out 
epoch in the early Universe. The freeze-out mechanism is based on the assumption that 
dark particles had been in thermal equilibrium in the early time at 
temperatures $T \gtrsim  m_{\text{DM}}$. In an expanding Universe the annihilation rate 
of dark particles into SM particles slows down and there 
is an epoch with $T \ll m_{\text{DM}}$ after which this rate descends 
below the Hubble expansion rate. On the other hand, from this time on dark particles 
are not kinematically allowed to get reproduced. Thus, in effect, the number density 
of dark particles, $n_{\chi}$, remains asymptotically constant within the comoving volume. 

The leading DM annihilation reactions which are necessary to determine the relic 
density are shown in Fig.~\ref{dark-ann}. In this figure the first and third 
annihilation processes occur via a $Z^\prime$ boson exchange 
in s-channel: $\chi \chi \to \bar b b, \bar t t, Z^\prime h, Z^\prime h^\prime$, while 
the second diagram shows annihilation with an intermediate 
DM via t- and u-channel: $\chi \chi \to Z^\prime Z^\prime$. 
The Boltzmann equation provides us with 
the evolution of DM number density in terms of thermal averaged annihilation 
cross sections $\langle \sigma_{\text{ann}}v_{\text{rel}} \rangle$ as
\ba
\frac{dn_{\chi}}{dt} +3Hn_{\chi} + \langle \sigma_{\text{ann}}v_{\text{rel}} \rangle [n^{2}_{\chi}-(n^{\text{EQ}}_{\chi})^2 ] = 0\,,
\ea
where, $n^{\text{EQ}}_{\chi}$ is the DM number density at 
equilibrium condition and $H$ is the Hubble parameter. 
In order to determine the present value of the number density and 
therefore the relic density one should solve numerically the Boltzmann equation
at freeze-out condition which is when the dark particles are away from equilibrium.     

We first implement our model into the program LanHEP \cite{Semenov:LanHEP} to give us all 
the basic vertices and Feynman rules of our model. Later on to analyze the DM 
relic density we employ the package MicrOMEGAs \cite{Belanger:MICRO} which requires our
output files from the LanHEP program. To check the validity of our model 
implementation into LanHEP, we utilize the program CalcHEP \cite{Belyaev:CalcHEP} using
our LanHEP outputs to calculate the annihilation cross sections. From this
we found agreements with our analytical calculations given in appendix \ref{formula} 
for the relevant annihilation cross sections.

\section{Numerical Analysis}
\label{Numerical}
\begin{figure}
\begin{minipage}{0.36\textwidth}
\includegraphics[width=\textwidth,angle =-90]{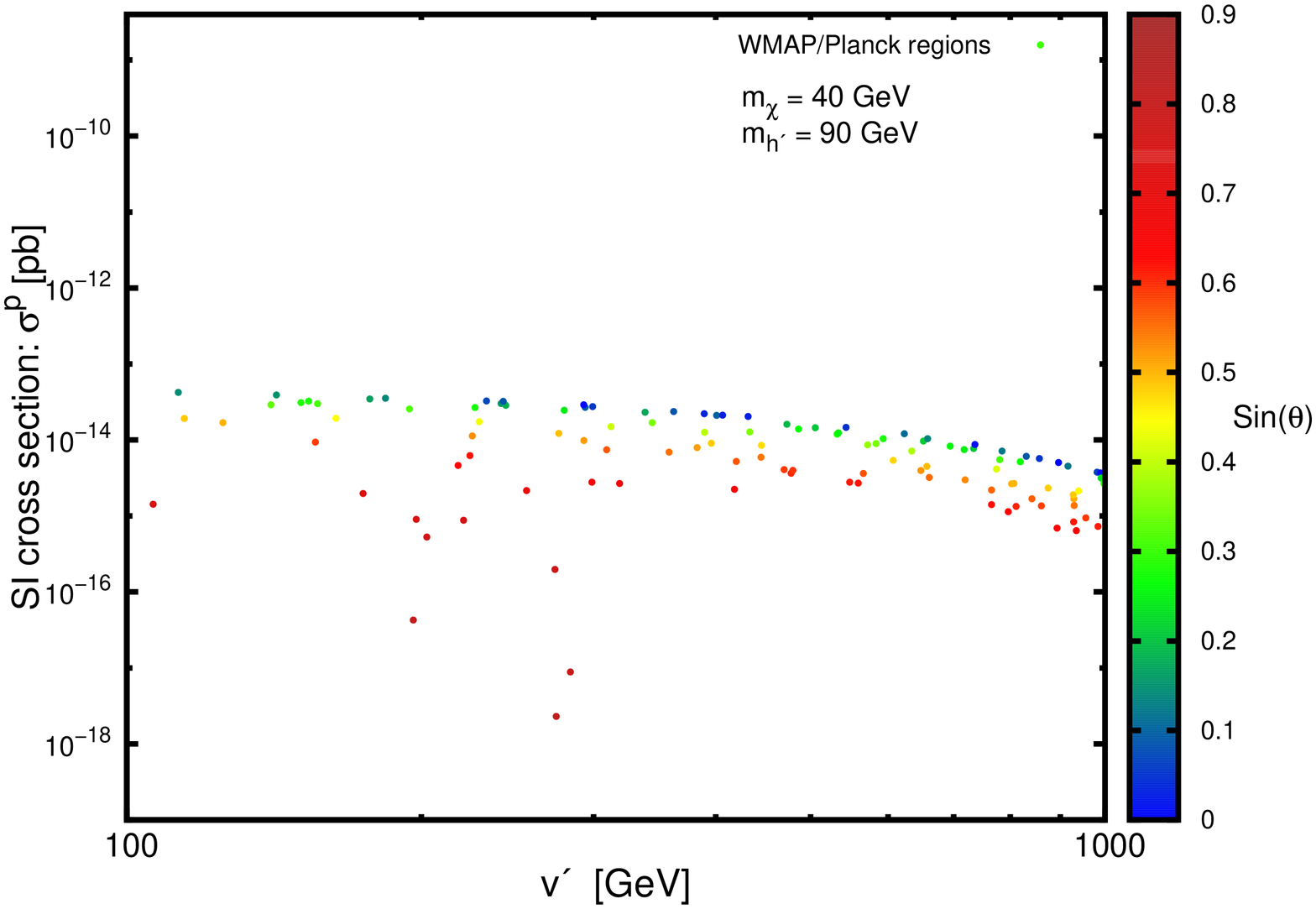}
\end{minipage}
\hspace{2.4cm}
\begin{minipage}{0.36\textwidth}
\includegraphics[width=\textwidth,angle =-90]{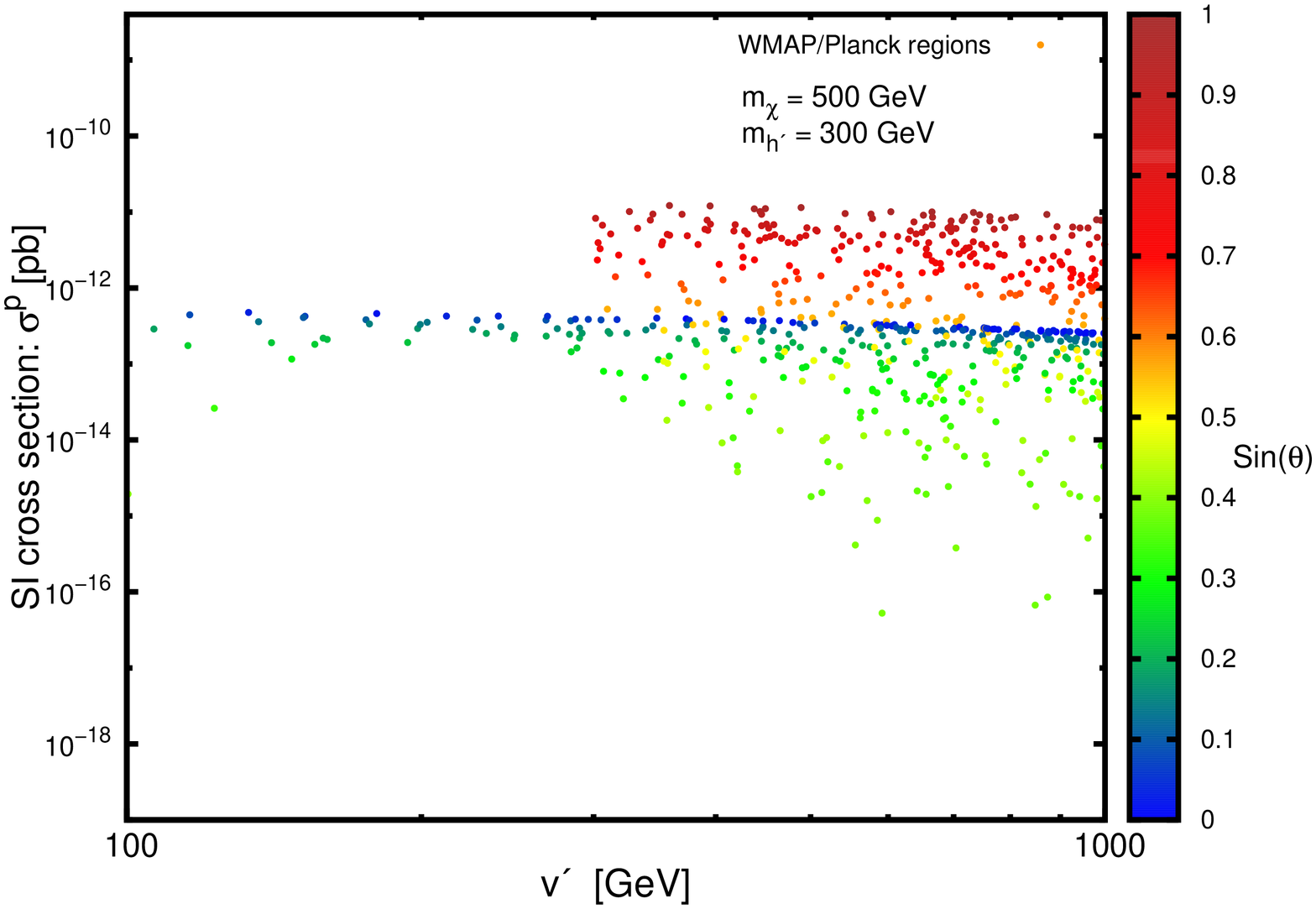}
\end{minipage}
\caption{Results for the spin-independent elastic scattering cross section of DM with proton 
as a function of two variables, scalar vacuum expectation value $v'$ and 
the mixing angle $\theta$. The coupling $g'$ is generated in the 
range $0.001 <  g' < 1$. For both values of the DM mass, the elastic scattering cross section
is below the upper limit given by LUX and XENON100.}
\label{vevscan}
\end{figure}

In this section we will find the viable region in the parameter space 
which respect observed relic density, invisible Higgs decay width measurement and 
constraints from direct detection experiments. We consider in our parameter space 
as independent free variables: $m_{\chi}$, $m_{h^\prime}$, $g^\prime$, $v^\prime$ and $\theta$. 
Throughout our study we keep fixed the SM Higgs mass as $m_{h} = 125$ GeV and 
the SM Higgs vacuum expectation value as $v = 246$ GeV. 
As a first numerical look, we would like to find the viable region in the parameter space 
for a given set of variables \{$m_{\chi},m_{h'}$\} by scanning over the scalar vacuum expectation value $v'$ 
and the mixing angle $\theta$.
To do so, we pick out random values for the coupling $g'$ in the reasonable range of $0.001 < g' < 1$.
Our results presented in Fig.~\ref{vevscan} for two sample values of the DM mass, 
indicate that it is possible to find viable regions in the parameter space for 
the wide range of $100 < v' < 1000$ and the mixing angle $0 < \theta < \pi/2$.    

To move on, we fix two variables out of five independent free parameters when we scan over 
the parameter space. For the vacuum expectation value of the singlet scalar 
we choose $v^\prime = 800$ GeV and perform our calculations for two different 
mixing angle with $\sin \theta = 0.01$ and $\sin \theta = 0.1$. 
It is then ensured that with these choices and the range of the masses we will 
pick out for the singlet scalar, the quartic couplings will respect bounds from 
perturbativity and vacuum stability conditions when relations in Eq.~(\ref{couplings}) 
are applied.  

We begin our scan over the parameter space by generating random values 
of order $\sim 10^5$  for three 
free parameters in the ranges:  $1$ GeV $< m_{\chi} < 1$ TeV, $20$ GeV $< m_{h^\prime} < 150$ GeV        
and $0.01 < g^\prime < 1$. Given the mass relation $m_{Z^\prime}  = g^\prime v^\prime /\sqrt{2}$,
the $Z^\prime$ boson mass will then lie in the range $5.6$ GeV $< Z^\prime < 565$ GeV.  
We then use the combined results from Planck \cite{Ade:2013zuv} and 
WMAP \cite{Hinshaw:2012aka}, $0.1172 < \Omega h^2 < 0.1226$, to 
exclude large regions in the parameter space which are 
inconsistent with these observations. 
At the same time, when $m_{Z^\prime}, m_{h^\prime} < m_{h}/2$ we check further
to make sure each generated point in the parameter space can fulfill the upper 
limit constraint on the invisible Higgs decay width.

\begin{figure}
\begin{minipage}{0.36\textwidth}
\includegraphics[width=\textwidth,angle =-90]{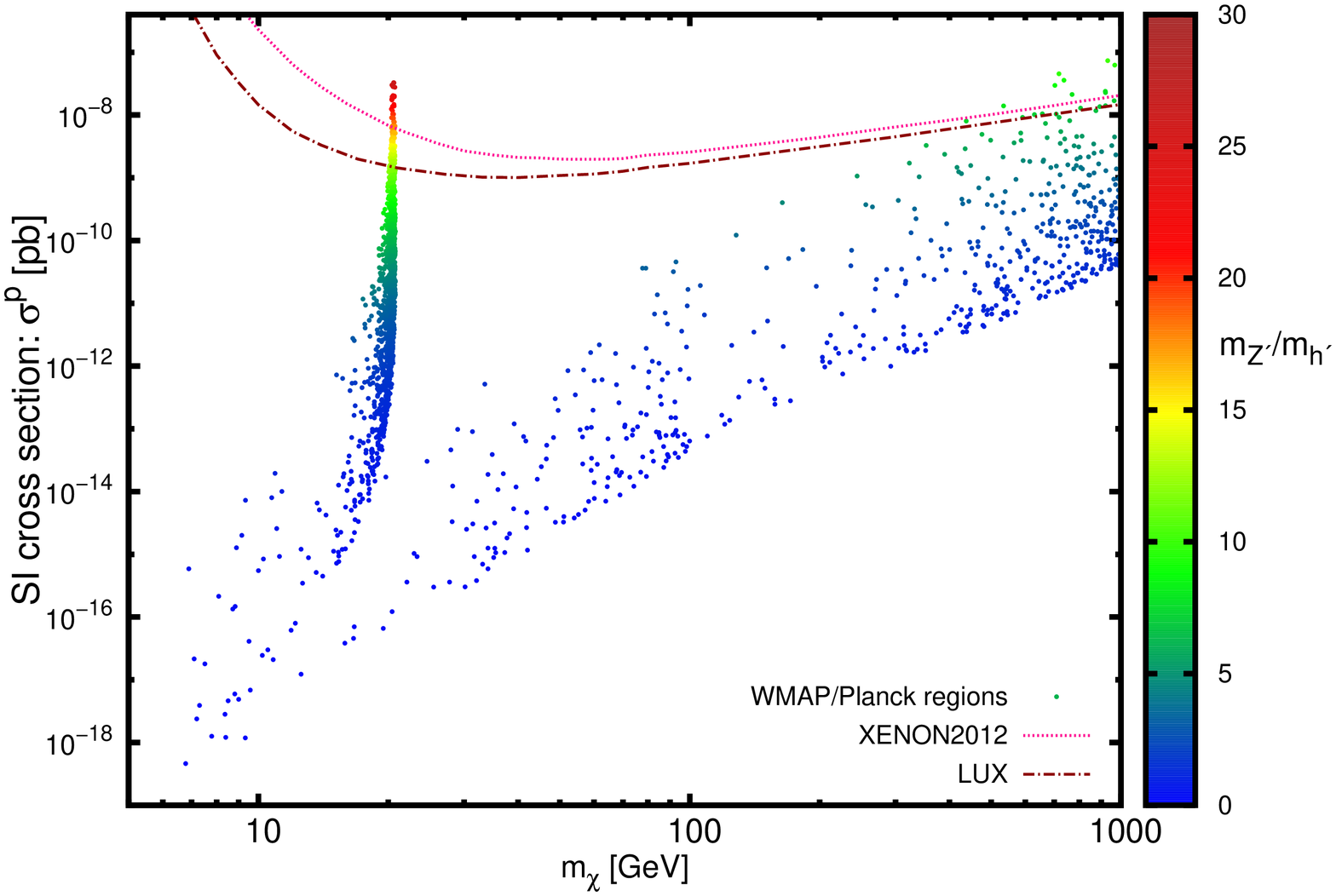}
\end{minipage}
\hspace{2.4cm}
\begin{minipage}{0.36\textwidth}
\includegraphics[width=\textwidth,angle =-90]{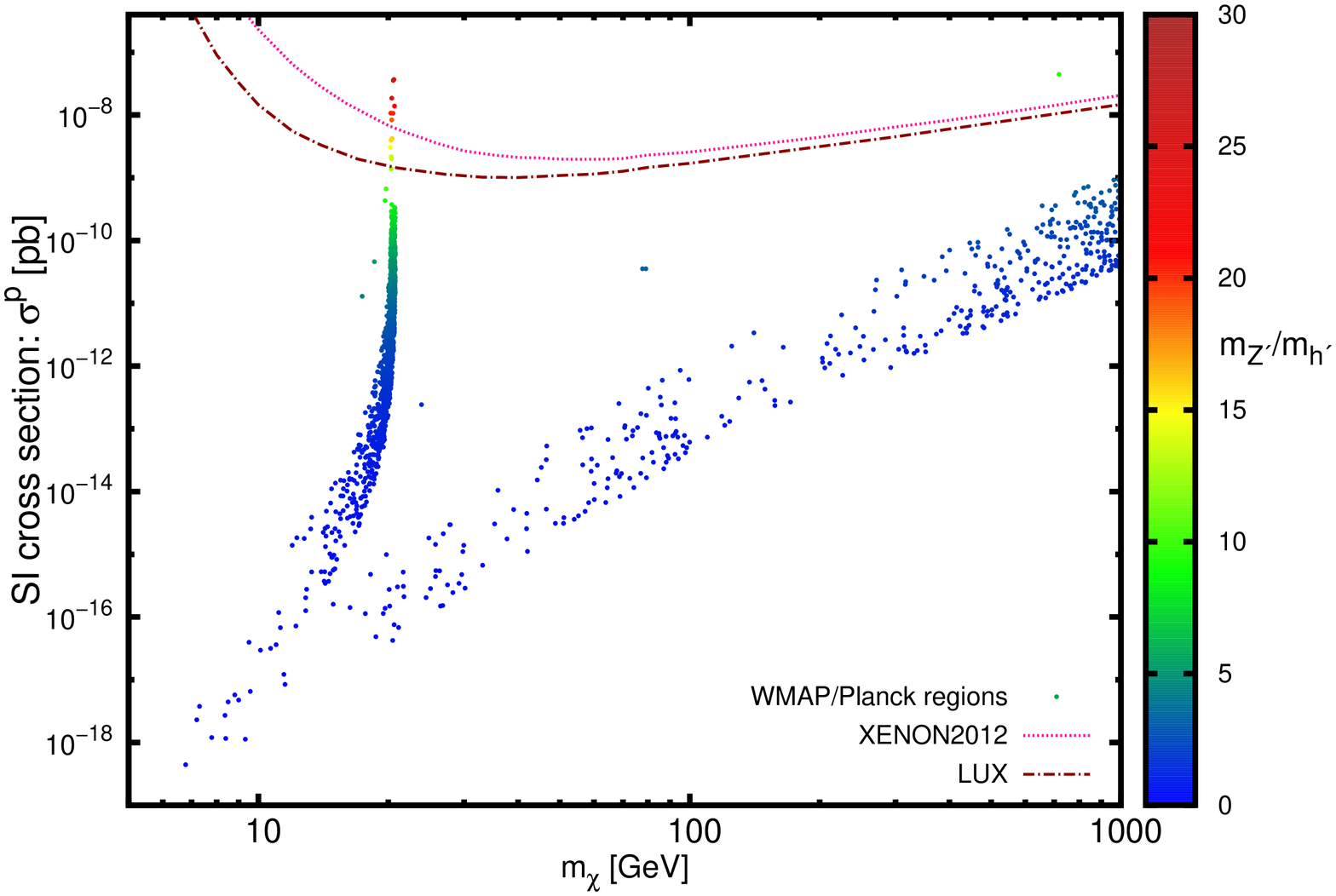}
\end{minipage}
\caption{Spin-independent elastic scattering cross section of DM with proton is shown
as a function of DM mass. All the points displayed in the plots respect constraints from 
observed relic density and invisible Higgs decay width measurement. 
The anticipated upper limit bounds on the elastic 
scattering cross section imposed by LUX and XENON100 experiments 
are placed to make comparison. The mixing angle is chosen $\sin \theta = 0.01$ in the left 
panel and $\sin \theta = 0.1$ in the right panel.}
\label{direct-analysis1}
\end{figure}

Using the formula provided by Eq.~(\ref{cross-section}) we compute the DM-proton 
elastic scattering cross section in terms of DM mass within 
the parameter space restricted by the observed relic density and invisible 
Higgs decay width measurement.  
For a wide range of the DM mass our results for the elastic 
scattering cross section are summarized in 
Fig.~\ref{direct-analysis1} for $\sin \theta = 0.01$ and $\sin \theta = 0.1$.

It is evident from these figures that in our model 
there exist viable regions in the parameter space 
with DM elastic scattering cross section well below LUX and 
XENON100 bounds when the ratio $m_{Z'}/m_{h'} \lesssim 5$.  
With $\sin \theta = 0.01$, all the points with correct relic abundance in the 
left panel of Fig.~\ref{direct-analysis1} respect the upper bound from the invisible 
Higgs decay. However, for larger mixing angle $\sin \theta = 0.1$, the invisible 
Higgs decay constraint excludes some portion of the region with correct relic 
density as can be seen by the right panel in Fig.~\ref{direct-analysis1}.  

Therefore we emphasize here on an interesting feature of 
the two-portal model that the dark particle can evade direct detection 
in the range of DM mass from a few GeV up to 1 TeV.

\section{Gamma-Ray Emission From DM Annihilation}
\label{gamma-ray-emission}
The gamma-ray excess observed in the GC from the analyses 
of the Fermi Large Area Telescope ({\it Fermi}-LAT) data is one of the 
places to look for the trace of the 
dark matter signals. Among other disfavored scenarios such as 
millisecond pulsars and cosmic-rays sources, the annihilation of the dark matter (which is more accumulated in the center 
of the Galaxy) into SM particles explains well the observed gamma-ray excess. 

After it was worked out in \cite{Hooper:2010mq} where the excess reported for the first time,
more accurate analyses were implemented by different groups confirming the original results 
\cite{Abazajian:2012pn,Macias:2013vya,Calore:2014xka}. 

In this section we examine the two-portal model discussed in the last sections for the 
gamma-ray excess. The region of interest (ROI) we use in our computation is that of
considered in \cite{Calore:2014xka}, i.e. at Galactic latitudes 
$ 2^\circ \leq \lvert b \lvert \leq 20^\circ$ and Galactic longitudes 
$\lvert l \vert \leq 20^\circ$ known as Inner Galaxy. 
\begin{figure}
\begin{center}
\includegraphics[scale=.5,angle =0]{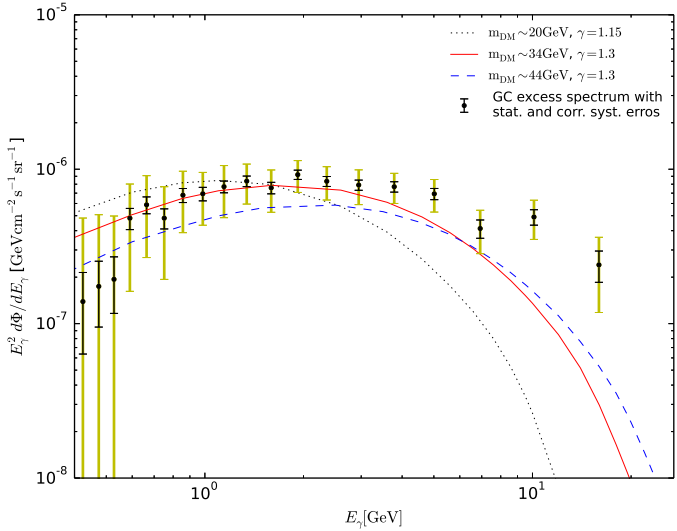}
\end{center}
\caption{The spectrum of the gamma-ray produced by DM annihilation for masses 
$m_{\text{DM}}=20, 34$ and $44$ GeV with slop parameter $\gamma= 1.15, 1,3$ and $1.3$ 
respectively for each DM mass. The DM mass $m_{\text{DM}}=34$ GeV is more compatible with
Fermi-LAT data analysis \cite{Calore:2014xka}.}
\label{flux}
\end{figure}

Let us briefly review the material we use to obtain the gamma-ray 
spectrum from dark matter annihilation.

The flux of the gamma-ray produced by annihilation of dark matter into SM particles is given by 

\begin{equation}
 \Phi(E_\gamma, \psi)=\frac{\braket{\sigma v}}{8\pi m_{\text{DM}}^2}\frac{dN_\gamma}{dE_\gamma} 
 \int_{\text{l.o.s}} \rho^2(r) dl\,,
\end{equation}
where $\braket{\sigma v}$  is the velocity averaged total annihilation cross section, 
$m_{\text{DM}}$ denotes the mass of the 
dark matter and $dN_\gamma /dE_\gamma$ is the gamma energy spectrum 
produced per annihilation. The integral of the density squared is performed over
the line-of-sight (l.o.s). The dark matter density as a function of $r$, the distance from the center of the Galaxy,
is given by $\rho(r)$. This density function is assumed to be spherically symmetric and 
is given by the generalized Navarro-Frenk-While (NFW) halo profile \cite{Navarro:1995iw,Navarro:1996gj},
\begin{equation}
 \rho(r)=\rho_0 \frac{(r/r_s)^{-\gamma}}{(1+r/r_s)^{3-\gamma}}\,,
\end{equation}
with the local dark matter density $\rho_0=0.4$ GeV/cm$^3$ and the radius scale $r_s=20$ kpc. 
Due to high uncertainty in the dark matter density near the center of our Milky Way galaxy, the inner slop parameter
takes values in the range $\gamma=1-1.3$.

We use micrOMEGAs to compute the gamma-ray spectrum for dark masses 
$m_{\text{DM}}\sim 20, 34, 44$ GeV that are picked out from the viable parameter space obtained
in the previous section (see Fig. \ref{direct-analysis1}). The gamma slop parameter is chosen 
$\gamma=1.15$ for $m_{\text{DM}}=20$ GeV and
$\gamma=1.3$ for $m_{\text{DM}}=34, 44$ GeV. The annihilation cross section that we 
obtain for different masses are $\braket{ \sigma v}= 2.14 \times 10^{-26} $ cm$^3$ s$^{-1}$ for 
$m_{\text{DM}}=20$ GeV,  $\braket{ \sigma v}= 2.41 \times 10^{-26} $ cm$^3$ s$^{-1}$ for 
$m_{\text{DM}}=34$ GeV and $\braket{ \sigma v}= 2.34 \times 10^{-26} $ cm$^3$ s$^{-1}$ for 
$m_{\text{DM}}=44$  GeV. In Fig. \ref{flux} we have plotted the energy 
spectrum of the gamma production for the masses mentioned above. As seen in this figure 
the DM mass $m_{\text{DM}}=34$ GeV has a better agreement with the {\it Fermi}-LAT data analysis
performed in \cite{Calore:2014xka}.

\section{Conclusions}
\label{sect:conclusions}

In this paper we have proposed a minimal fermionic dark matter model with two  
Higgs and vector portals in tandem, both charged under a $U(1)'$ symmetry. 
The $Z'$ dark boson beside coupling to the fermionic DM
has leptophobic interaction with the SM particles. The $Z'$ coupling to the light quarks
has also been considered negligible, resulting in a suppressed DM-nucleon interaction. The 
leading contribution to the DM-nucleon elastic scattering comes from a one-loop Feynman diagram
as shown in Fig.~\ref{dark-direct}. 

An interesting result is that we find a wide range of DM mass from a few GeV up to $1$ TeV  
in the viable parameter space respecting all constraints from observed relic
abundance, direct detection bounds and invisible Higgs decay width. 

The dark matter annihilation into SM particles results in the gamma-ray energy spectrum that fits best
with the {\it Fermi}-LAT data for the DM mass $m_{\text{DM}}=34$ GeV.\\

The new aspects of the two-portal DM in comparison with the earlier works 
which involve either the Higgs portal or the $Z'$ portal are discussed below.\\

1) In the fermionic DM model introduced in \cite{Kim:2008pp}, DM interacts with the SM particles 
   with exchanging the SM Higgs ($h$) or a new scalar particle ($h'$). They found out that 
   almost the entire parameter space is excluded by the XENON and CDMS except a very small 
   resonant regions corresponding to very restricted conditions for the DM mass: 
   $m_{\chi} \sim m_{h}/2$ or $m_{\chi} \sim m_{h'}/2$. 
   It is likely that with future experiments like XENON1T, 
   these small viable regions get even smaller. 
   Moreover, the model discussed in \cite{Kim:2008pp} predicts a large invisible Higgs decay width, 
   such that with the current bound from the LHC, the DM masses less than about 50 GeV 
   are excluded.   
   In contrast, the presented model in this paper with simultaneous two portals  
   provides us with a rather wide viable parameter space well below XENON100 and LUX bounds. 
   More precisely, we see that the viable range for the DM mass is from a few GeV up to 1 TeV.
   It is quite unlikely that the future proposed or planned direct detection experiments can 
   rule out the entire viable parameter space for the aforementioned range of the DM mass.  
   One more interesting feature of the proposed model is that the total Higgs decay width
   including the effects of the new particles does not depend on the DM mass. Therefore, 
   even for small DM mass it becomes possible to find viable regions in the parameter space 
   which respect the invisible Higgs decay width bound.   \\
            
  2) In the two-portal model, the coupling between the fermionic DM 
     and the $Z'$ boson is of vector type while the anomaly free condition imposes 
     a pure axial coupling between the $Z'$ boson with the third family of quarks. 
     One crucial feature of the present model is the fact that with the help of scalar 
     mediator (Higgs portal) and the non-universality of the $Z'$ boson, 
     spin-independent elastic scattering of the DM with nuclei is possible. 
     In the earlier works (see e.g. \cite{Berlin:2014tja,Hooper:2014fda}) with the same axial coupling 
     between $Z'$ boson and the third family of quarks there is no DM-nucleon elastic scattering at all.
     Therefore we have shown in the two-portal model that even without having any $Z'$ interaction with 
     $u$ and $d$ quarks, there is a prospect for a signal at the direct detection experiments.


 \appendix
 \section{Dark Matter Annihilation Cross Sections}
 \label{formula}
We give the DM annihilation cross section formulas in this section. 
The annihilation cross section for the process $\chi \chi \to \bar f f$ with $f = b, t$ is obtained as
\ba
\label{ff}
\sigma_{\text{ann}} v_{\text{rel}} (\bar \chi \chi \to \bar f f) = 
\frac{{g^\prime}^4 \sqrt{1-4m_{f}^2/s}}{2 \pi s} 
\Big[ \frac{s^2-8m_{f}^2 m_{\chi}^2 + 2 s m_{\chi}^2 -\frac{4}{9}s m_{f}^2}{(s-m^{2}_{Z'})^2+m^{2}_{Z'}\Gamma^{2}_{Z'}}
 \Big] \,.
\ea
The next annihilation process $\chi \chi \to h Z'$, is again mediated 
by a dark gauge boson via s-channel. We get the following result for the cross section as,  
\ba
\label{z'h}
\sigma_{\text{ann}} v_{\text{rel}} (\bar \chi \chi \to h Z') = 
\frac{{g^\prime}^6 v'^2 \sin^2 \theta}{16 \pi s} \frac{(s+2 m_{\chi}^2) \sqrt{[1-(m_{h}^2+m_{Z'}^2)/s]^2-4m_{h}^2m_{Z'}^2/s^2}}
{(s-m^{2}_{Z'})^2+m^{2}_{Z'}\Gamma^{2}_{Z'}} \,.
\ea
We then obtain the DM annihilation cross section for the process $\chi \chi \to h' Z'$, 
\ba
\label{z'h'}
\sigma_{\text{ann}} v_{\text{rel}} (\bar \chi \chi \to h' Z') = 
\frac{{g^\prime}^6 v'^2 \cos^2 \theta}{16 \pi s} \frac{(s+2 m_{\chi}^2) \sqrt{[1-(m_{h'}^2+m_{Z'}^2)/s]^2-4m_{h'}^2m_{Z'}^2/s^2}}
{(s-m^{2}_{Z'})^2+m^{2}_{Z'}\Gamma^{2}_{Z'}} \,.
\ea
Finally, we get the annihilation cross section for the process $\chi \chi \to Z' Z'$ which 
takes place by mediating a DM via t- and u-channel,  
\ba
\label{Z'Z'}
\sigma_{\text{ann}} v_{\text{rel}} (\bar \chi \chi \to Z' Z') =
\frac{{g^\prime}^4 \sqrt{1-4m_{Z'}^2/s}}{8 \pi^2 s}
\int d\Omega \Big[ \frac{s m_{Z'}^2 - m_{\chi}^2m_{Z'}^2 + \frac{1}{2} s m_{\chi}^2-2m_{\chi}^4}{(t-m_{\chi}^2)(u-m_{\chi}^2)}
\nonumber \\
- \frac{(m_{\chi}^2+m_{Z'}^2-t)^2+ts -s m_{\chi}^2+2 t m_{\chi}^2+4 m_{\chi}^2 m_{Z'}^2+2 m_{\chi}^4 }
{2(t-m_{\chi}^2)^2}
\nonumber \\
- \frac{(m_{\chi}^2+m_{Z'}^2-u)^2+us -s m_{\chi}^2+2 u m_{\chi}^2+4 m_{\chi}^2 m_{Z'}^2+2 m_{\chi}^4 }
{2(u-m_{\chi}^2)^2} \Big] \,,
\ea
where, $s$, $t$ and $u$ are the relevant mandelstam variables.


\begin{thebibliography}{99}

\bibitem{Ade:2013zuv}
  P.~A.~R.~Ade {\it et al.}  [Planck Collaboration],
  Astron.\ Astrophys.\  {\bf 571} (2014) A16
  [arXiv:1303.5076 [astro-ph.CO]].

\bibitem{Hinshaw:2012aka}
  G.~Hinshaw {\it et al.}  [WMAP Collaboration],
  Astrophys.\ J.\ Suppl.\  {\bf 208} (2013) 19
  [arXiv:1212.5226 [astro-ph.CO]].


\bibitem{Akerib:LUX}
  D.~S.~Akerib {\it et al.}  [LUX Collaboration],
  Phys.\ Rev.\ Lett.\  {\bf 112} (2014) 091303
  [arXiv:1310.8214 [astro-ph.CO]].

\bibitem{Aprile:XENON100}
  E.~Aprile {\it et al.}  [XENON100 Collaboration],
  Phys.\ Rev.\ Lett.\  {\bf 109} (2012) 181301
  [arXiv:1207.5988 [astro-ph.CO]].

\bibitem{Hooper:2010mq}
  D.~Hooper and L.~Goodenough,
  Phys.\ Lett.\ B {\bf 697} (2011) 412
  [arXiv:1010.2752 [hep-ph]].

\bibitem{Abazajian:2012pn}
  K.~N.~Abazajian and M.~Kaplinghat,
  Phys.\ Rev.\ D {\bf 86} (2012) 083511


\bibitem{Gordon:2013vta}
  C.~Gordon and O.~Macias,
  Phys.\ Rev.\ D {\bf 88} (2013) 083521
  [arXiv:1306.5725 [astro-ph.HE]].

\bibitem{Abazajian:2014fta}
  K.~N.~Abazajian, N.~Canac, S.~Horiuchi and M.~Kaplinghat,
  Phys.\ Rev.\ D {\bf 90} (2014) 023526
  [arXiv:1402.4090 [astro-ph.HE]].

\bibitem{Daylan:2014rsa}
  T.~Daylan, D.~P.~Finkbeiner, D.~Hooper, T.~Linden, S.~K.~N.~Portillo, N.~L.~Rodd and T.~R.~Slatyer,
  arXiv:1402.6703 [astro-ph.HE].

\bibitem{Calore:2014xka}
  F.~Calore, I.~Cholis and C.~Weniger,
  arXiv:1409.0042 [astro-ph.CO].

\bibitem{Calore:2014nla}
  F.~Calore, I.~Cholis, C.~McCabe and C.~Weniger,
  Phys.\ Rev.\ D {\bf 91} (2015) 6,  063003
  [arXiv:1411.4647 [hep-ph]].

\bibitem{Boehm:2014hva}
  C.~Boehm, M.~J.~Dolan, C.~McCabe, M.~Spannowsky and C.~J.~Wallace,
  JCAP {\bf 1405} (2014) 009
  [arXiv:1401.6458 [hep-ph]].

\bibitem{Ipek:2014gua}
  S.~Ipek, D.~McKeen and A.~E.~Nelson,
  Phys.\ Rev.\ D {\bf 90} (2014) 5,  055021
  [arXiv:1404.3716 [hep-ph]].

\bibitem{Izaguirre:2014vva}
  E.~Izaguirre, G.~Krnjaic and B.~Shuve,
  Phys.\ Rev.\ D {\bf 90} (2014) 5,  055002
  [arXiv:1404.2018 [hep-ph]].

\bibitem{Cheung:2014lqa}
  C.~Cheung, M.~Papucci, D.~Sanford, N.~R.~Shah and K.~M.~Zurek,
  Phys.\ Rev.\ D {\bf 90} (2014) 7,  075011
  [arXiv:1406.6372 [hep-ph]].

\bibitem{Ghorbani:2014qpa}
  K.~Ghorbani,
  JCAP {\bf 1501} (2015) 015
  [arXiv:1408.4929 [hep-ph]].
  
  
 \bibitem{Dolan:2014ska}
  M.~J.~Dolan, C.~McCabe, F.~Kahlhoefer and K.~Schmidt-Hoberg,
  JHEP {\bf 1503} (2015) 171
  [arXiv:1412.5174 [hep-ph]].


\bibitem{Gherghetta:2015ysa}
  T.~Gherghetta, B.~von Harling, A.~D.~Medina, M.~A.~Schmidt and T.~Trott,
  arXiv:1502.07173 [hep-ph].


\bibitem{Dudas:2009uq}
  E.~Dudas, Y.~Mambrini, S.~Pokorski and A.~Romagnoni,
  JHEP {\bf 0908} (2009) 014
  [arXiv:0904.1745 [hep-ph]].

\bibitem{Chu:2013jja}
  X.~Chu, Y.~Mambrini, J.~Quevillon and B.~Zaldivar,
  JCAP {\bf 1401} (2014) 01,  034
  [arXiv:1306.4677 [hep-ph]].

\bibitem{Arcadi:2013qia}
  G.~Arcadi, Y.~Mambrini, M.~H.~G.~Tytgat and B.~Zaldivar,
  JHEP {\bf 1403} (2014) 134
  [arXiv:1401.0221 [hep-ph]].


\bibitem{Lebedev:2014bba}
  O.~Lebedev and Y.~Mambrini,
  Phys.\ Lett.\ B {\bf 734} (2014) 350
  [arXiv:1403.4837 [hep-ph]].

\bibitem{Belanger:2007dx}
  G.~Belanger, A.~Pukhov and G.~Servant,
  JCAP {\bf 0801} (2008) 009
  [arXiv:0706.0526 [hep-ph]].

\bibitem{Buckley:2011mm}
  M.~R.~Buckley, D.~Hooper and J.~L.~Rosner,
  Phys.\ Lett.\ B {\bf 703} (2011) 343
  [arXiv:1106.3583 [hep-ph]].

\bibitem{Pospelov:2007mp}
  M.~Pospelov, A.~Ritz and M.~B.~Voloshin,
  Phys.\ Lett.\ B {\bf 662} (2008) 53
  [arXiv:0711.4866 [hep-ph]].

\bibitem{Hooper:2014fda}
  D.~Hooper,
  Phys.\ Rev.\ D {\bf 91} (2015) 3,  035025
  [arXiv:1411.4079 [hep-ph]].

\bibitem{Berlin:2014tja}
  A.~Berlin, D.~Hooper and S.~D.~McDermott,
  Phys.\ Rev.\ D {\bf 89} (2014) 11,  115022
  [arXiv:1404.0022 [hep-ph]].
  
\bibitem{Mizukoshi:2010ky}
  J.~K.~Mizukoshi, C.~A.~de S.Pires, F.~S.~Queiroz and P.~S.~Rodrigues da Silva,
  Phys.\ Rev.\ D {\bf 83} (2011) 065024
  [arXiv:1010.4097 [hep-ph]].


\bibitem{Alves:2015pea}
  A.~Alves, A.~Berlin, S.~Profumo and F.~S.~Queiroz,
  arXiv:1501.03490 [hep-ph].



\bibitem{London:1986}
D.~London and J.~L.~Rosner, Phys.\ Rev.\ D {\bf 34}, 1530
(1986).


\bibitem{Hewett:1988xc}
  J.~L.~Hewett and T.~G.~Rizzo,
  Phys.\ Rept.\  {\bf 183} (1989) 193.



\bibitem{Langacker:2008yv}
  P.~Langacker,
  Rev.\ Mod.\ Phys.\  {\bf 81} (2009) 1199
  [arXiv:0801.1345 [hep-ph]].

\bibitem{Hill:2002ap}
  C.~T.~Hill and E.~H.~Simmons,
  Phys.\ Rept.\  {\bf 381} (2003) 235
   [Phys.\ Rept.\  {\bf 390} (2004) 553]
  [hep-ph/0203079].

\bibitem{delAguila:1986iw}
  F.~del Aguila, G.~A.~Blair, M.~Daniel and G.~G.~Ross,
  Nucl.\ Phys.\ B {\bf 283} (1987) 50.

\bibitem{ArkaniHamed:2001is}
  N.~Arkani-Hamed, A.~G.~Cohen and H.~Georgi,
  Phys.\ Lett.\ B {\bf 516} (2001) 395
  [hep-th/0103135].

\bibitem{ArkaniHamed:2002qx}
  N.~Arkani-Hamed, A.~G.~Cohen, E.~Katz, A.~E.~Nelson, T.~Gregoire and J.~G.~Wacker,
  JHEP {\bf 0208} (2002) 021
  [hep-ph/0206020].

\bibitem{Han:2003wu}
  T.~Han, H.~E.~Logan, B.~McElrath and L.~T.~Wang,
  Phys.\ Rev.\ D {\bf 67} (2003) 095004
  [hep-ph/0301040].

\bibitem{Perelstein:2005ka}
  M.~Perelstein,
  Prog.\ Part.\ Nucl.\ Phys.\  {\bf 58} (2007) 247
  [hep-ph/0512128].



\bibitem{Chacko:2005pe}
  Z.~Chacko, H.~S.~Goh and R.~Harnik,
  Phys.\ Rev.\ Lett.\  {\bf 96} (2006) 231802
  [hep-ph/0506256].


\bibitem{Demir:2005ti}
  D.~A.~Demir, G.~L.~Kane and T.~T.~Wang,
  Phys.\ Rev.\ D {\bf 72} (2005) 015012
  [hep-ph/0503290].


  
\bibitem{Ghorbani:2014gka}
  K.~Ghorbani and H.~Ghorbani,
  arXiv:1501.00206 [hep-ph].
    
\bibitem{Cline:2015qha}
  J.~M.~Cline, G.~Dupuis, Z.~Liu and W.~Xue,
  arXiv:1503.08213 [hep-ph].
  


\bibitem{Hill:1994hp}
  C.~T.~Hill,
  Phys.\ Lett.\ B {\bf 345} (1995) 483
  [hep-ph/9411426].


\bibitem{Andrianov:1998hx}
  A.~A.~Andrianov, P.~Osland, A.~A.~Pankov, N.~V.~Romanenko and J.~Sirkka,
  Phys.\ Rev.\ D {\bf 58} (1998) 075001
  [hep-ph/9804389].



\bibitem{Chivukula:2002ry}
  R.~S.~Chivukula and E.~H.~Simmons,
  Phys.\ Rev.\ D {\bf 66} (2002) 015006
  [hep-ph/0205064].







\bibitem{Liu:2011dh}
  J.~Y.~Liu, Y.~Tang and Y.~L.~Wu,
  J.\ Phys.\ G {\bf 39} (2012) 055003
  [arXiv:1108.5012 [hep-ph]].
  
\bibitem{Heinemeyer:2013}
  S.~Heinemeyer {\it et al.}  [LHC Higgs Cross Section Working Group Collaboration],
  arXiv:1307.1347 [hep-ph].

\bibitem{Belanger:invisible}
  G.~Belanger, B.~Dumont, U.~Ellwanger, J.~F.~Gunion and S.~Kraml,
  Phys.\ Lett.\ B {\bf 723} (2013) 340
  [arXiv:1302.5694 [hep-ph]].


\bibitem{Belanger:2008-Direct}
  G.~Belanger, F.~Boudjema, A.~Pukhov and A.~Semenov,
  Comput.\ Phys.\ Commun.\  {\bf 180} (2009) 747
  [arXiv:0803.2360 [hep-ph]].


\bibitem{Ellis:2008}
  J.~R.~Ellis, K.~A.~Olive and C.~Savage,
  Phys.\ Rev.\ D {\bf 77} (2008) 065026
  [arXiv:0801.3656 [hep-ph]].

\bibitem{Nihei:2004}
  T.~Nihei and M.~Sasagawa,
  Phys.\ Rev.\ D {\bf 70} (2004) 055011
   [Erratum-ibid.\ D {\bf 70} (2004) 079901]
  [hep-ph/0404100].

\bibitem{Belanger:MICRO}
  G.~Belanger, F.~Boudjema, A.~Pukhov and A.~Semenov,
  Comput.\ Phys.\ Commun.\  {\bf 185} (2014) 960
  [arXiv:1305.0237 [hep-ph]].
  
  
\bibitem{Semenov:LanHEP}
  A.~Semenov,
  arXiv:1005.1909 [hep-ph].

\

\bibitem{Belyaev:CalcHEP}
  A.~Belyaev, N.~D.~Christensen and A.~Pukhov,
  Comput.\ Phys.\ Commun.\  {\bf 184} (2013) 1729
  [arXiv:1207.6082 [hep-ph]].


\bibitem{Macias:2013vya}
  O.~Macias and C.~Gordon,
  Phys.\ Rev.\ D {\bf 89} (2014) 6,  063515
  [arXiv:1312.6671 [astro-ph.HE]].

\bibitem{Navarro:1995iw}
  J.~F.~Navarro, C.~S.~Frenk and S.~D.~M.~White,
  Astrophys.\ J.\  {\bf 462} (1996) 563
  [astro-ph/9508025].


\bibitem{Navarro:1996gj}
  J.~F.~Navarro, C.~S.~Frenk and S.~D.~M.~White,
  Astrophys.\ J.\  {\bf 490} (1997) 493
  [astro-ph/9611107].

\bibitem{Kim:2008pp}
  Y.~G.~Kim, K.~Y.~Lee and S.~Shin,
  JHEP {\bf 0805} (2008) 100
  [arXiv:0803.2932 [hep-ph]].


\end{thebibliography}
\end{document}